\documentclass[12pt]{iopart}
\usepackage{graphicx}

\begin{document}

\title[Creating thermal distributions from diabatic \ldots]{Creating thermal distributions from diabatic excitations
in ion-trap-based quantum simulation}

\author{M. H. Lim}

\address{Department of Physics, Harvard University, 17 Oxford Street, Cambridge, Massachusetts 02138 USA}

\author{B. T. Yoshimura}

\address{Department of Physics, Georgetown University, 37$^{\rm th}$ and O Sts. NW, Washington, District of Columbia 20057 USA}

\author{ J. K. Freericks}

\address{Department of Physics, Georgetown University, 37$^{\rm th}$ and O Sts. NW, Washington, District of Columbia 20057 USA}

\ead{james.freericks@georgetown.edu}
\vspace{10pt}
\begin{indented}
\item[]
\date{today}
\end{indented}

\begin{abstract}
The goal of adiabatic ground-state preparation is to start a simple quantum system in its ground state and adiabatically
evolve the Hamiltonian to a complex one, maintaining the ground state throughout the evolution. In ion-trap-based quantum simulations, coherence times are too short to allow for adiabatic evolution for large chains, so the system evolves diabatically, creating excitations to higher energy states. Because the probability for diabatic excitation depends exponentially on the excitation energy and because the thermal distribution also depends exponentially on the excitation energy, we investigate whether the diabatic excitation can create a thermal distribution; as this could serve as an alternative for creating thermal states of complex quantum systems without requiring contact with a heat bath. In this work, we explore this relationship and determine situations where diabatic
excitation can approximately create such a thermal state.
\end{abstract}

%
\vspace{2pc}

%
\submitto{\NJP}
%
%
%

\section{Introduction}

Quantum simulation in ion traps has made significant progress over the past five years. Initial work examined the 
transverse-field Ising model with adiabatic state preparation~\cite{kim,islam1,islam2} and more recently with excited state spectroscopy (since the experiments created significant diabatic excitations)~\cite{senko,yoshimura,roos}. The latest
work has examined Lieb-Robinson bounds in Ising and XY systems~\cite{blatt,richerme} and higher-spin variants~\cite{richerme2}. In general, the coherence time of the experiments is too short to allow for adiabatic state preparation (which becomes worse for frustrated spin systems, and for larger numbers of spins), and the system has significant diabatic excitation. In this work, we examine how close the eigenstate populations from a diabatic excitation are to those of a thermal distribution and we compare common experimental measurements (like the spin structure factor or the Binder cumulant) between these two cases as well.

In the controlled quantum environment of an isolated quantum system, one would like to be able to examine
situations that represent equilibrium physical behavior.  Namely, we would like to be able to create thermal 
distributions of a complex quantum system from some simple initial states of well-understood quantum systems. This approach goes a step beyond conventional adiabatic state preparation because its goal is not just to create the 
ground state, but to create the appropriate mixture of quantum states that corresponds to a thermal distribution without having to place the system in the contact with a thermal reservoir. 
If this goal can be achieved, then one would have a highly tunable quantum simulator that can solve many equilibrium quantum problems and can determine critical phenomena like scaling exponents and other universal properties of phase transitions in the presence of large quantum effects. At the moment, we do not have such flexibility in conventional simulations, but because the energy gap determines the excited state populations both in the diabatic case and in the
thermal case, it seems possible that these two cases may have situations where the diabatic excitations do appear to be thermal. We investigate this possibility here.

The system that we work with is a transverse-field Ising model with static Ising exchange parameters and a time-dependent transverse field. The Ising exchange parameters are long-range and approximately decay with a power law in the inter-ion distance that
is tunable between the uniform case ($\alpha=0$) and the dipole-dipole case ($\alpha=3$). They are calculated exactly
following the results in Ref.~\cite{duan_monroe}, and they approximately satisfy
\begin{equation}\label{eq: jij}
|J_{ij}| \approx\frac{J_0}{|\frac{R_i}{a}-\frac{R_j}{a}|^{\alpha}},
\end{equation}
with $J_0$ an overall scale for the exchange interactions, $R_i$ the position of the $i^{\rm th}$ ion and $a$ the average nearest-neighbor distance between ions in the chain (hence the denominator is nearly $|i-j|^\alpha$). The equilibrium positions of the ions in a harmonic trap are not uniformly spaced, but they are nearly so for the chains we consider in this work.
All $J_{ij}$ are positive for the ferromagnetic case and negative for the antiferromagnetic case. The Hamiltonian
becomes
\begin{equation}
\mathcal{H}(t)= -\sum_{\scriptsize\begin{array}{c}
i,j=1\\
i<j
\end{array}}^N J_{ij}\sigma_z^{(i)}\sigma_z^{(j)}-B_{x}(t)\sum_{i=1}^N\sigma_x^{(i)}.
\label{eq: ham}
\end{equation}
Here, $\sigma_\gamma^{(i)}$ is the Pauli spin matrix (with eigenvalues $\pm 1$ and with $\gamma=x$, $y$, or $z$ denoting the spatial direction of the Pauli matrix) at lattice site $i$, $B_x(t)$ is
the time-dependent transverse field, and $N$ is the number of spins in the lattice; we work in units with $\hbar=1$
and simulate the transverse-field Ising model in a linear Paul trap.  This model is generated in the ion trap by applying an optical spin-dependent force (which employs two optical beams with slightly different frequencies), and integrating out the effects of the 
phonons (assuming they are only virtually occupied)~\cite{duan_monroe}. In general, the spin exchange parameters are time-dependent, but in experiments with the detuning of the difference of the two optical beams to the blue of the transverse center-of-mass mode---where all of the exchange coefficients have the same sign---the system is approximated well by the static spin exchange parameters~\cite{wang_freericks}, even if phonons are excited during the drive from the laser.

\section{Methods}

The explicit formula for the static spin exchange coefficients is~\cite{duan_monroe}
\begin{equation}
J_{ij}=\Omega^2\nu_R\sum_{m=1}^N\frac{b_{im}b_{jm}}{\mu^2-\omega_m^2},
\end{equation}
and we use the experimental parameters from Ref.~\cite{islam2} where $\Omega= 600$~kHz is the Rabi frequency, $\nu_R=h/(M\lambda^2)=18.5$~kHz is the recoil energy of a $^{171}{\rm Yb}^+$ ion (with $h$ Planck's constant, $M$ the mass of the ion, and $\lambda=355$~nm the wavelength of the laser light), $b_{im}$ is the value of the orthonormal eigenvector at the $i^{\rm th}$ ion site of the $m^{\rm th}$ transverse normal mode for the $N$-ion chain, $\omega_m$ is the corresponding normal mode frequency, and $\mu=\omega_{\rm COM}+3\eta\Omega=1.0233\omega_{\rm COM}$ is the detuning from the transverse center of mass mode (with $\eta=\sqrt{\nu_R/\omega_{\rm COM}}=0.0621$ the Lamb-Dicke parameter). We work in conventional frequency units throughout. The range of the spin exchange coefficients is adjusted by adjusting the anisotropy between the transverse and the axial traps, keeping the transverse trap fixed. The axial center of mass mode then has a frequency running from 620~kHz to 950~kHz, corresponding to a nearest neighbor exchange interaction which is near 1~kHz ($J_0\approx 1$~kHz) and the power law decay running from $0.7<\alpha<1.2$.

The protocol we follow for the adiabatic state preparation (with diabatic excitations) is as follows: (i) initialize the system with all spins in the $x$ direction and with the field large and positive $B_x\gg J_0$, so it starts in the ground state and (ii) reduce the magnetic field in an exponential fashion with $B_x(t)=B_0\exp(-t/\tau)$ for a time constant $\tau$ with $J_0\tau\approx 1/2$ and the total time interval for the evolution being on the order of $6\tau$ time units; the initial magnetic field satisfies $B_0=5J_0$. With $J_0\approx 1$~kHz, we have $\tau \approx 0.5$~ms and the total running time for the experiment being 3~ms, similar to the
experimental run times of recent experiments. We also choose the trap asymmetry so that the power law for the decay of the spin exchange satisfies $0.5\le \alpha\le 2$.  We work with chains ranging in size from $N=6$ to $N=12$.

The time evolution of the system is calculated by evolving the wavefunction forward in time. We employ the Crank-Nicolson~\cite{crank} algorithm to do this, with a small enough step size in the time discretization that unitarity
of the time evolution is preserved throughout the simulation. This is checked explicitly by reducing the step size until results do not change within the precision of the calculation.

We will be comparing the diabatic evolution of the ground-state wavefunction to the mixed state of a thermal
distribution.  In order to do this, we need to identify a strategy for determining the effective temperature of the thermal distribution for this comparison.  Of course, if the system evolved fully into a thermal distribution, then all of the different techniques we use to identify the effective temperature would agree.  But because the evolution is not exact, these different strategies can yield different results. We summarize these strategies next.

The first thing we have to realize is that one difference between the diabatic evolution and a thermal distribution is that the diabatic evolution can only populate quantum states that have the same symmetry as the initial ground state.  For the transverse-field Ising model, with long-range couplings, there are two symmetries that arise.  The first is a spatial reflection symmetry, which can be expressed as $J_{ij}=J_{N-iN-j}$, if we number the lattice sites in the chain from left to right in increasing order. This symmetry arises from the fact that the ion positions in the trap  have a reflection symmetry about the origin in the axial direction, and so do the normal modes (due to the even symmetry of the trapping potentials). The second symmetry is a spin-reflection parity, where we perform the unitary transformation $\sigma_x\rightarrow \tilde\sigma_x$, $\sigma_y\rightarrow -\tilde\sigma_y$ and $\sigma_z\rightarrow -\tilde\sigma_z$, which leaves the Hamiltonian and the spin-spin commutation relations invariant. Both of these reflection symmetries produce eigenvalues with respect to the parity of the reflection, which can be even or odd for the spatial and for the spin symmetry, separately. Hence, the diabatic evolution will only populate states in the initial symmetry sector of the ground state; the other symmetry sectors are unchanged due to the diabatic evolution, and remain unpopulated. Note that this last statement can be relaxed if phonons are actually created during the time evolution, rather than just being virtually created.  This is because the spin-reflection parity is not a symmetry of the laser-ion interacting Hamiltonian, but only of the effective spin Hamiltonian. 

We employ three different strategies to extract an effective temperature for the thermal distribution that we will use to compare to the diabatic distribution. The first one is to find the effective temperature of the thermal distribution that has the same average energy as that of the time-evolved state. If we define the (orthonormal) eigenstates of the system to satisfy $\mathcal{H}(t_f)|n\rangle=E_n|n\rangle$ at the end of the experiment, then the partition function becomes $\mathcal{Z}=\sum_n \exp(-\beta E_n)$, with $\beta=1/T$ the
inverse temperature (setting $k_B=1$) and $t_f$ the final time for the evolution of the system. Then, if we denote the final time-evolved wavefunction by $|\psi(t_f)\rangle$, the effective temperature for the average energy fit solves
\begin{equation}
\langle E\rangle_{\rm dia}=\langle \psi(t_f)|\mathcal{H}(t_f)|\psi(t_f)\rangle = \langle E\rangle_{\rm therm}=\frac{1}{\mathcal{Z}}\sum_n e^{-\beta E_n} E_n,
\label{eq: ave_energy_therm}
\end{equation}
where the sum goes over all of the eigenstates of the final Hamiltonian.
The only adjustable parameter is $\beta$, which is adjusted to solve the above equation. This method is called the thermal average fit for the effective temperature.
The second method we use is to find the effective temperature that has the same thermal fluctuations of the energy about the mean. This relation becomes
\begin{eqnarray}
&~&\langle (\Delta E)^2\rangle_{\rm dia}=\langle \psi(t_f)|\mathcal{H}^2(t_f)|\psi(t_f)\rangle - \langle E\rangle^2_{\rm dia}\nonumber\\
&=&\langle (\Delta E)^2\rangle_{\rm therm}=\frac{1}{\mathcal{Z}}\sum_n e^{-\beta E_n}E_n^2-\langle E\rangle_{\rm therm}^2,
\label{eq: fluc_energy_therm}
\end{eqnarray}
which also has just one parameter to adjust for the fit---$\beta$. This method is called the thermal fluctuation fit for the effective temperature.
The third strategy is to relate the ratio of the probability to be in the first excited state over the probability to be in the ground state to the Boltzmann formula for that ratio in terms of the excitation energy. Namely we use
\begin{equation}
e^{-\beta (E_1-E_{\rm gs})}=\frac{P_1}{P_{gs}},
\end{equation}
or
\begin{equation}
\beta = \frac{\ln P_{\rm gs}-\ln P_1}{E_1-E_{\rm gs}},
\end{equation}
where $P_n=|\langle n |\psi(t_f)\rangle|^2$. This last result is only meaningful when the probability to be in the ground state is larger than the probability to be in the first excited state, otherwise it produces a negative effective temperature, which is not a thermally stable state. This method is called the thermal ratio fit for the effective temperature.

\section{Results}

We examined the diabatic evolution of the transverse-field Ising model, employing parameters similar to those used in experiment, for a range of different systems including $N=6$, 8, 10, and 12, $\alpha$ ranging from 0.5 to 2 in steps of 0.25, and for the ferromagnetic $J_0>0$ and antiferromagnetic $J_0<0$ cases. In general, we found that the effective temperature was determined best by the thermal average fit. The fit from the thermal ratio often would yield negative effective temperatures for the antiferromagnetic case, while the fit for the thermal fluctuations tended to produce too high an effective temperature to properly fit the lower excited states. In cases where all three fits are close to one another (which occurred more often for the ferromagnetic case), we can often infer that the system is nearly in a thermal state.

\begin{figure}
\centering
\includegraphics[scale=0.18]{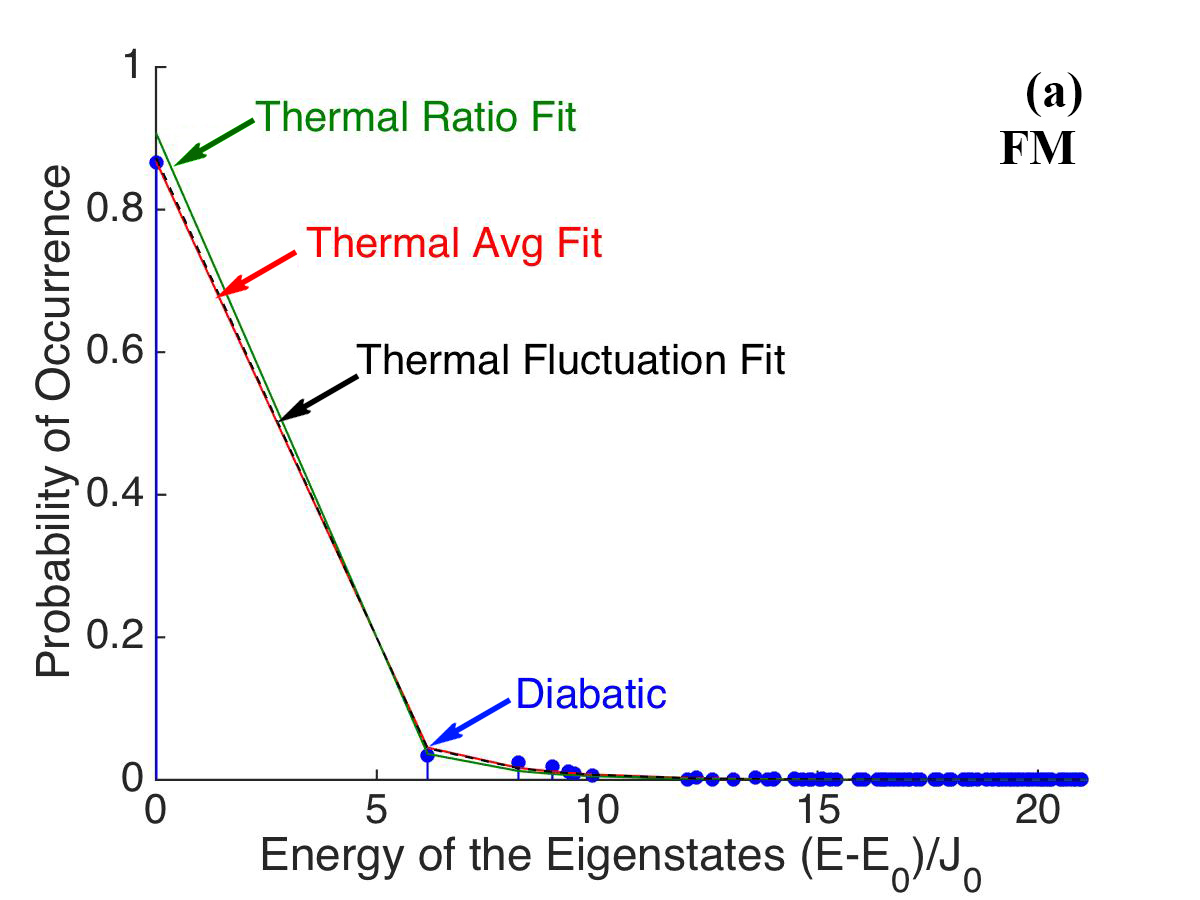}\includegraphics[scale=0.18]{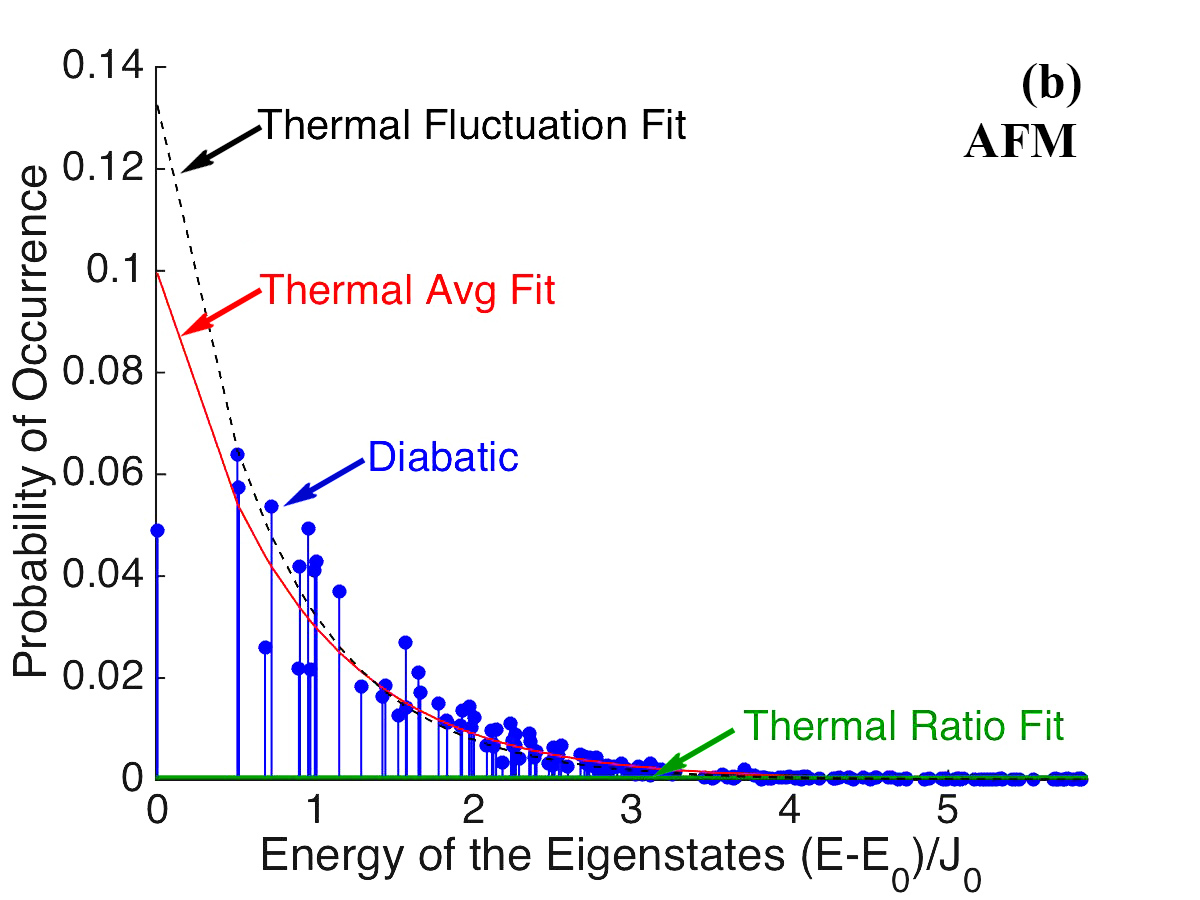}
\caption{Probability distribution of energy eigenstates in the same symmetry sector as the ground state for $N=10, \alpha=1$, when the system is diabatically evolved for $6\tau=t_f=3.0$~ms. Panel (a) on the left is for the ferromagnetic case, while panel (b) on the right is for the antiferromagnetic case. The effective temperature fit using the average energy usually yields the best fit compared to the other choices.}
\label{fig:probability}
\end{figure}

Figure~\ref{fig:probability} shows the comparison of the probability distribution for the energy eigenstates at time $t_f$ for the diabatically evolved state to those of a thermal distribution with an effective temperature fit. The ferromagnetic case matches well to the thermal distribution, while the antiferromagnetic case does not. Note that the antiferromagnetic case usually has the ground-state probability lower than the first excited-state probability except in cases of large $\alpha$ when the Ising coupling becomes closer to a nearest-neighbor interaction. While the ferromagnetic case is described well by a thermal distribution. 

One interesting feature arises in the antiferromagnetic case. The data for the probabilities appear to fit distributions with two distinct effective temperatures, one linked to the ground-state probability and one to the first excited-state probability. This behavior can be clearly seen in Fig.~\ref{fig:probability}~(b). While we do not have any conclusive explanation for why this behavior occurs, we did notice that the states in each of the distributions tend to be characterized by an expansion in terms of two localized states or four localized states when one simultaneously diagonalizes the energy, spatial parity, and spin parity. If we apply the parity symmetry operations (spatial and spin) to a given localized state, it either maps back onto the state itself, or it maps onto another state, so if we group localized basis states in the $z$-axis orientation according to their spatial and spin parity eigenstates, these states include two or four terms in the expansion; they also become eigenstates when $B_x=0$. 

\begin{figure}
\centering
\includegraphics[scale=0.20]{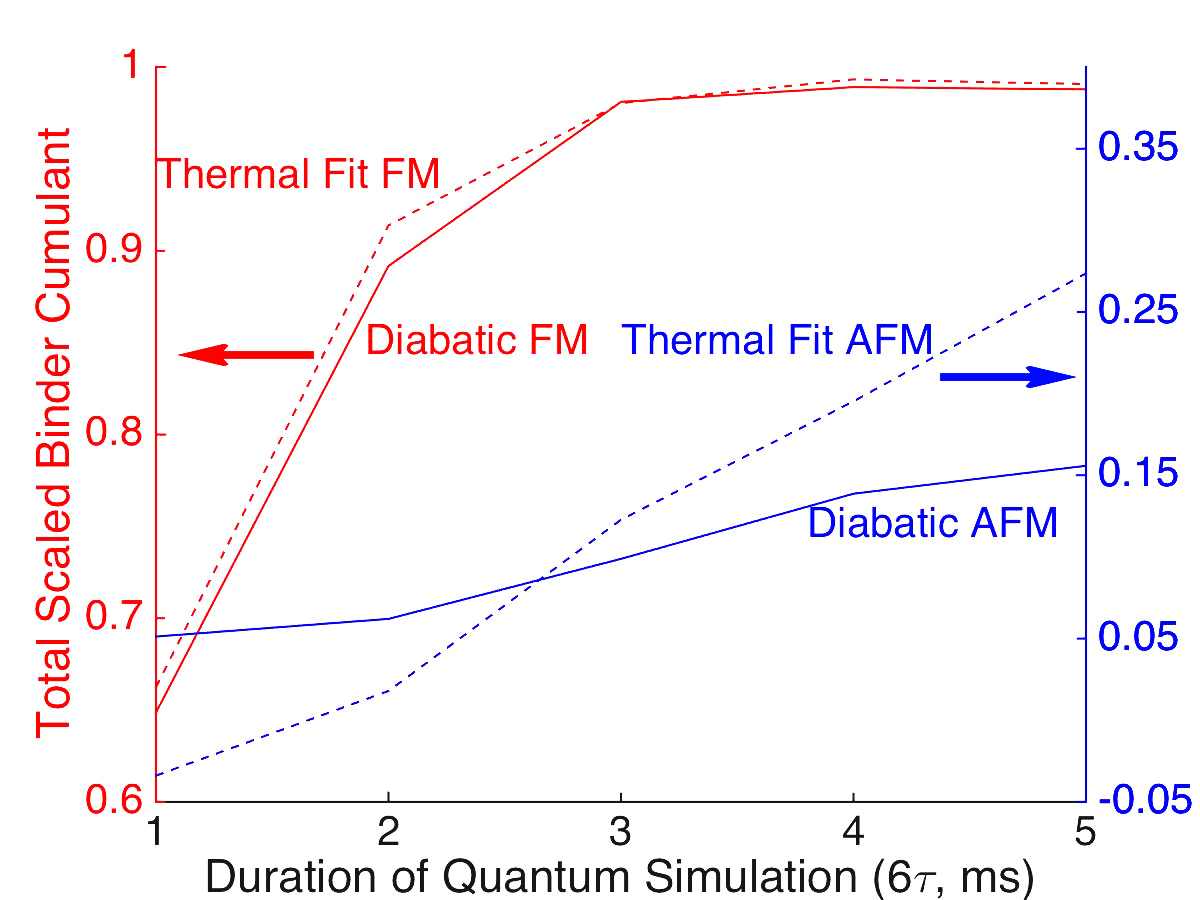}
\caption{Binder cumulant for the diabatic evolution (solid line) compared to a thermal distribution with an effective temperature given by the thermal average fit (dashed line) for $N=10$ and $\alpha=1$ as the simulation time $t_f$ is increased from 1~ms to 5~ms (in all cases $\tau=t_f/6$). The diabatic results are given by solid lines and the thermal fits by dashed lines. The FM case (red) uses the scale on the left, while the AFM case (blue) uses the scale on the right.}
\label{fig:binder}
\end{figure}

We measure the expectation values of various operators that are typically measured in an experiment.  This allows additional perspective for comparing the diabatic state to the thermal distribution. One common operator is the Binder cumulant~\cite{binder}, which is defined by
\begin{equation}\label{eq:binder}
g_s = \frac{\langle{(m_s-\langle{m_s}\rangle)^4}\rangle}{\langle{(m_s-\langle{m_s}\rangle)^2}\rangle^2}.
\end{equation}
Here, the operator $m_s$ stands for the uniform magnetization operator for the ferromagnetic case and the staggered magnetization operator for antiferromagnetic case: $m_s=\displaystyle\frac{1}{N}\sum_{i=1}^{N}(\pm1)^i \sigma_z^{(i)}$. We further scale the calculated Binder cumulants to $\bar{g_s}=(g_s^0-g_s)/(g_s^0-1)$ with $g_s^0=3-2/N$ to remove finite-size effects and to have the Binder cumulant vary from 0 in the least-ordered state to 1 in the most-ordered state. Fig.~\ref{fig:binder} shows these results.

The ferromagnet rapidly orders as we decrease the ramping speed and lengthen the simulation time. The ferromagnetic thermal fit produces a similar Binder cumulant, indicating that the diabatic Binder cumulant can be described with a corresponding thermal Binder cumulant. On the other hand, the antiferromagnet does not become strongly ordered as we lengthen the simulation time. Moreover, the corresponding thermal fit for the antiferromagnet has a Binder cumulant that is less than zero for small values of $6\tau$ and is larger than the diabatic result when the simulation is done for longer times. As we increase the number of ions $N$ and decrease the power law exponent $\alpha$, the Binder cumulant becomes smaller and takes a longer time to achieve order. Nevertheless, the trends remain the same. Once again, we find that the antiferromagnetic case is poorly approximated by a thermal distribution.

\begin{figure}
\centering
\includegraphics[scale=0.18]{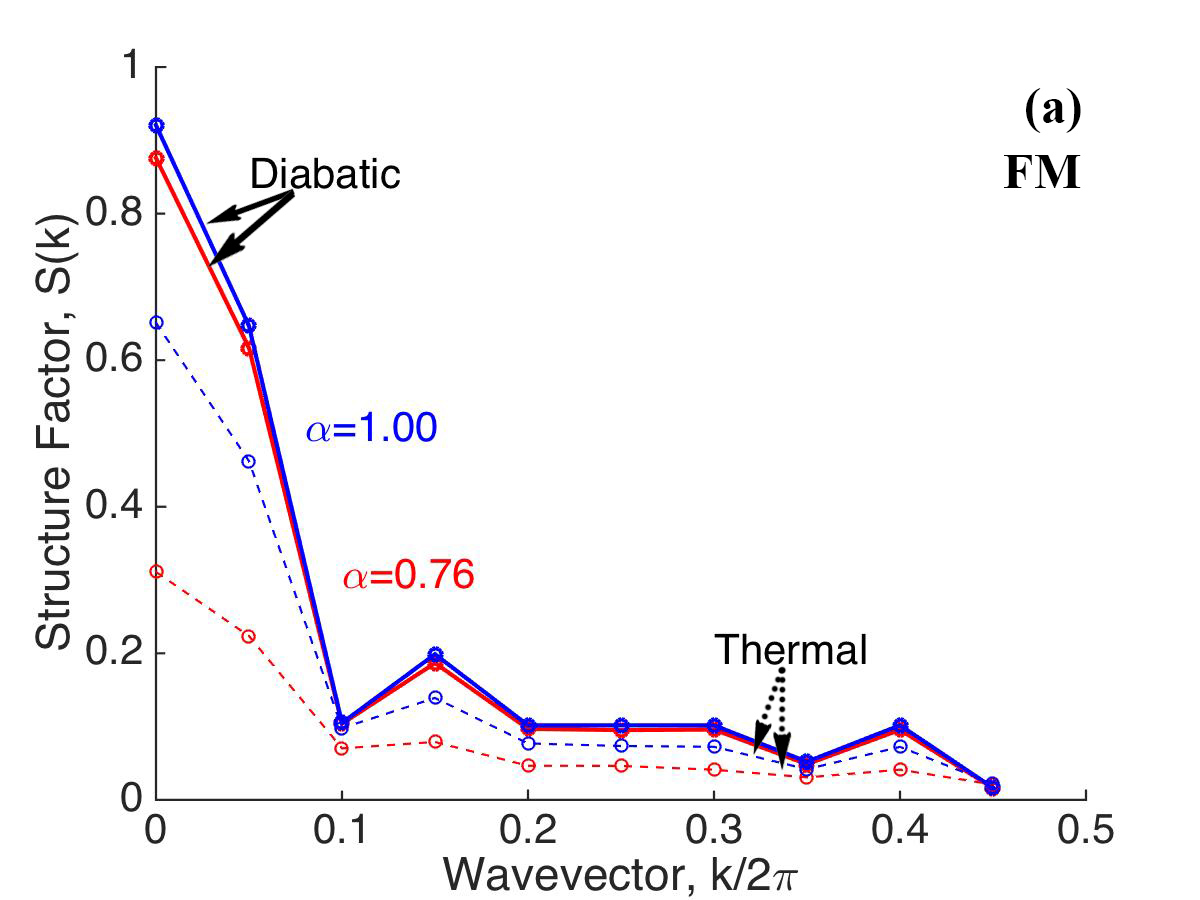}\includegraphics[scale=0.18]{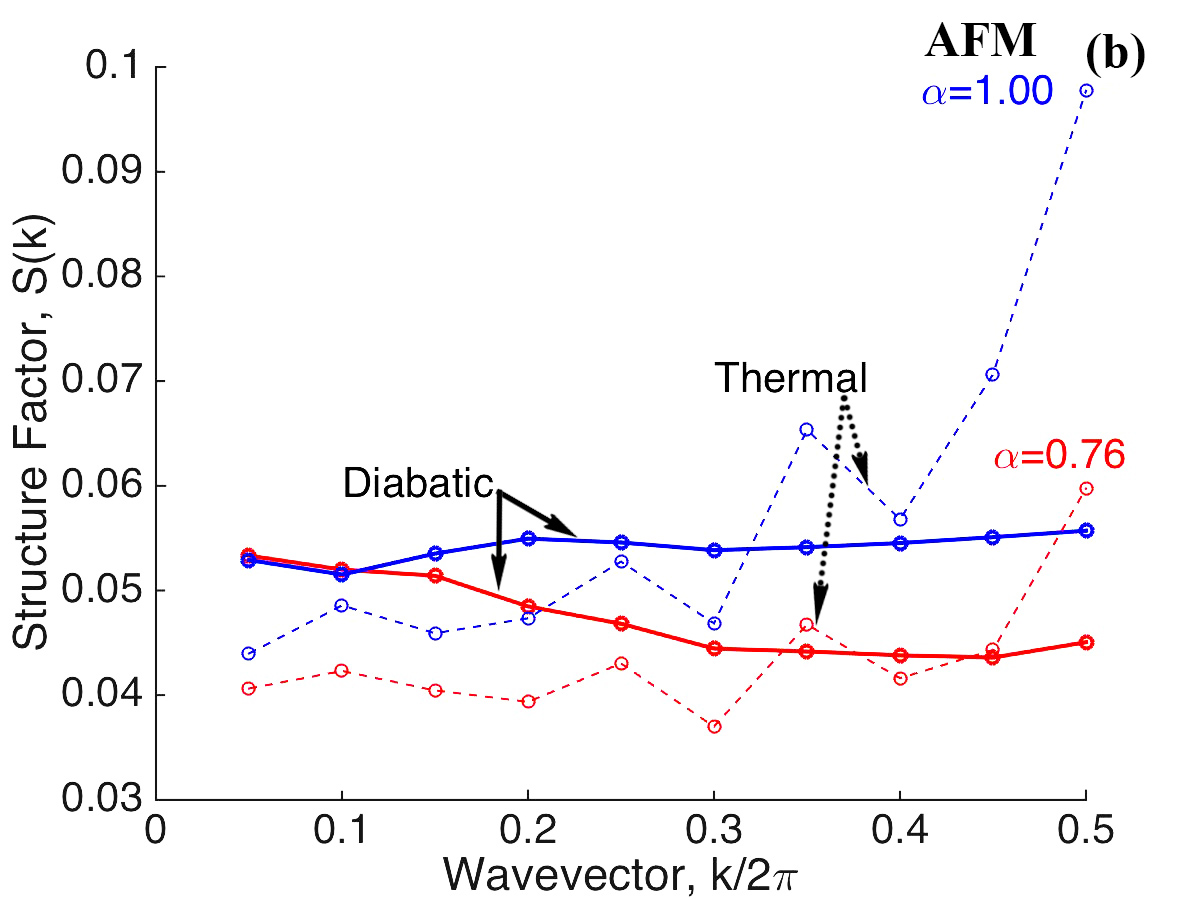}
\caption{ Structure factor for the diabatically evolved state (solid line) compared to the thermal distribution (dashed line) for $N=10$ and $6\tau=3$~ms.}
\label{fig:structure}
\end{figure}

Another operator that we calculated is the magnetic structure factor. The magnetic structure factor is the Fourier transform of the static spin-spin correlation function $C_{i,j} = \langle{\sigma_z^{(i)}\sigma_z^{(j)}}\rangle-\langle{\sigma_z^{(i)}}\rangle\langle{\sigma_z^{(j)}}\rangle$, which measures the correlation between two spins at sites $i$ and $j$. The formula for the structure factor becomes
\begin{equation}\label{eq:structure}
S(k) = \displaystyle\frac{1}{N-1}\left | \sum_{r=1}^NC(r)e^{ikr}\right |.
\end{equation}
Here, $C(r)=\displaystyle\frac{1}{N-r}\sum_{m=1}^{N-r}C_{m,m+r}$ is the average correlation between spins separated by $r$ sites, and $k$ is the wave number ($-\pi\le k \le \pi$). The larger the structure factor is, the more ordered the spin system is for a spin distortion that is modulated by the wavevector $k$.  Figure~\ref{fig:structure} plots the structure factor for two different exponents $\alpha=0.76$ and $1$.

For the ferromagnetic case, the structure factor shows a close agreement between the diabatically evolved state and the thermal distribution. As $\alpha$ is increased---corresponding to shorter-ranged spin-spin couplings---the structure factor for the time-evolved system becomes closer to the thermal distribution result. Of course, these results are peaked near $k=0$, since that is where the ferromagnetic order is the strongest.  The opposite holds for the antiferromagnetic case.  Increasing $\alpha$ (reducing the range of he interaction) produced worse agreement between the time-evolved and the thermal states. Overall, the ordering is reduced (and is now peaked around the $k=\pi$ value).

\begin{figure}[htb]
\centering
\includegraphics[scale=0.18]{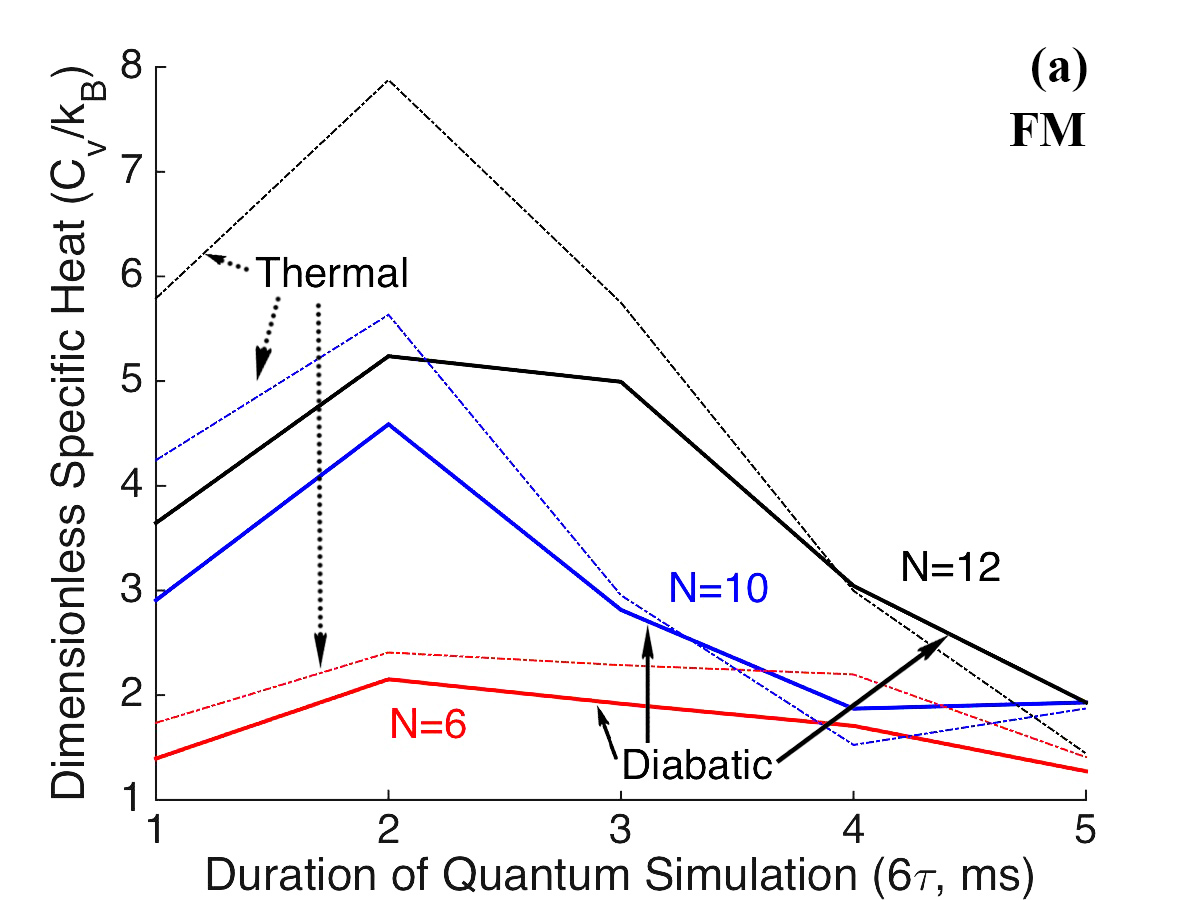}\includegraphics[scale=0.18]{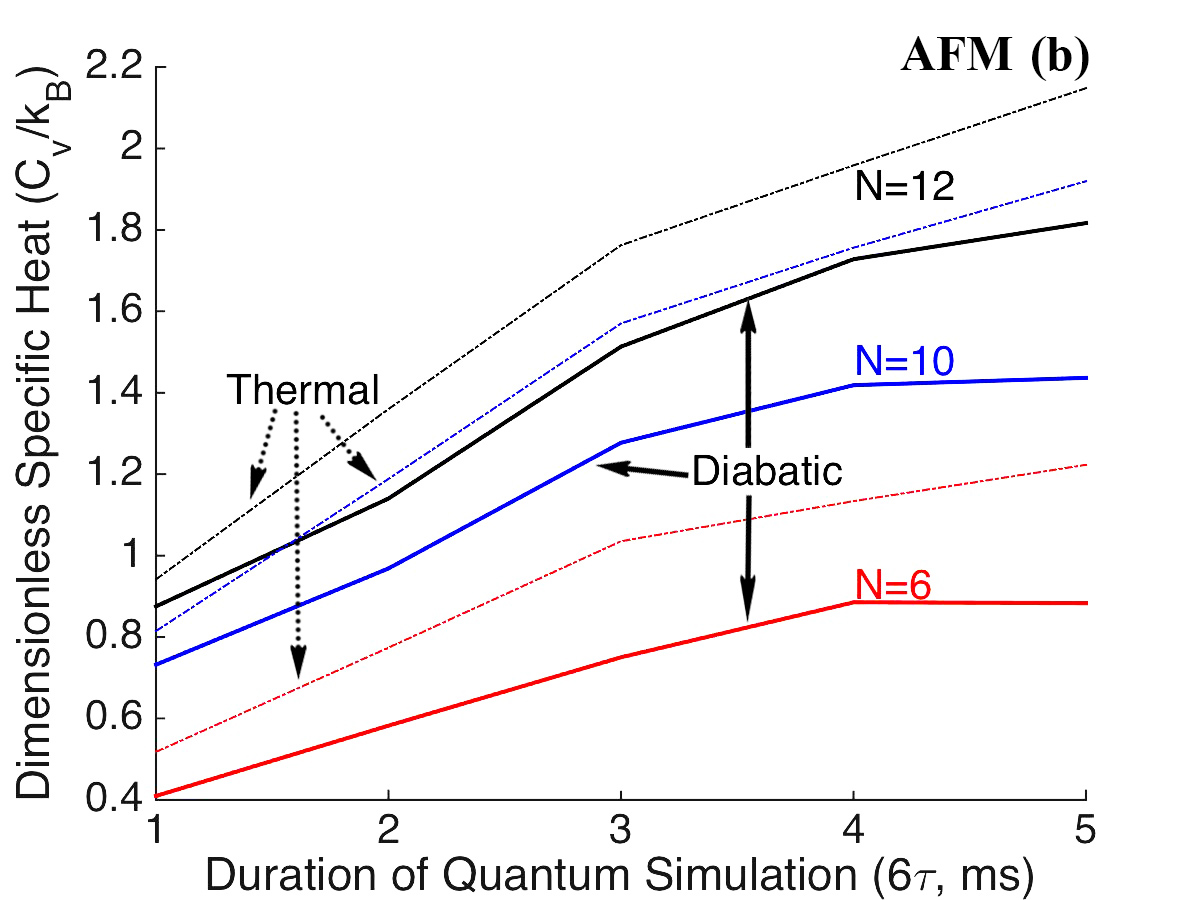}
\caption{Generalized specific heat for the diabatic state (solid line) compared to the equilibrium specific heat fit with the thermal average fit (dotted line).  The parameters are $\alpha=1$, $6\tau=t_f$, $N=6$, 8, 10, and 12, and $t_f$ ranging from 1~ms to 5~ms.}
\label{fig:heat}
\end{figure}

We also examine an analog of the specific heat. The traditional definition of the specific heat holds only in equilibrium, because the definition involves the temperature of the system and is given by
\begin{equation}\label{eq:heat}
C_v^{\rm therm} = \frac{\partial \langle{E}\rangle}{\partial T}= \frac{\langle (\Delta E)^2\rangle_{\rm therm}}{T^2}.
\end{equation} 
Here, we generalize the definition for the diabatic case, by employing the effective temperature that is fit with the thermal average fit via $C_v^{\rm dia}=\langle (\Delta E)^2\rangle_{\rm dia}/T^2_{\rm eff}$. These results are plotted in Fig.~\ref{fig:heat} for $N=6$, 8, 10, and 12.

The ferromagnetic specific heat has an interesting peak that develops near $t_f=2$~ms. It is unclear what the origin of this peak is. We do see, however, that the time-evolved diabatic state and the thermal distribution results agree well for long enough experimental times, but the agreement gets worse as $N$ increases. The antiferromagnetic case has a fairly flat specific heat and worse agreement between the diabatic and the equilibrium results, but the shape is preserved well for different lengths of the chains and when we compare the diabatic result with the equilibrium result. It looks like the specific heat is not a good indicator of the differences between the thermal and the diabatic cases.

\section{Discussion}

The transverse-field Ising model is a workhorse model for testing many different quantum theories given its position as one of the simplest models with a nontrivial phase transition. It is employed in a wide range of analog quantum simulators, particularly those that use ultracold ions in traps. One of the proposed modes of operation for a quantum simulator is 
adiabatic state preparation, where the quantum system is prepared in a simple state and the Hamiltonian is modified, adiabatically, to evolve the system into the ground state of a nontrivial Hamiltonian. It turns out this goal is difficult to achieve in most quantum simulators, especially as the size of the system grows and as the gap to excited states gets smaller. So, an interesting corollary would be to create thermal states in the system without requiring the system to be attached to a bath (which could lead to decoherence and other problems). We have performed a series of calculations to test this idea, examining both ferromagnetic and antiferromagnetic cases. We find that the idea seems to work quite well for ferromagnetic cases. Three different ways that one can extract an effective temperature from the data tend to agree, and the value of the effective temperature is governed by the rate that the Hamiltonian is changed, which determines the extent of the diabatic excitations created in the system. We compared a number of different observables as well, and found that generically, they also seem to be reproduced accurately by this approach. So, while it is clear that there are some differences, the idea does seem to work well under one caveat. Namely, the diabatic excitation only excites states that are directly coupled to the original ground state.  If the system has quantum numbers that are preserved during the time-dependent evolution of the system, then they will not allow for diabatic excitation into any other symmetry sector. Perhaps the equilibrium results are still accurately produced because of the eigenstate thermalization hypothesis~\cite{eigenstate}, which says that the expectation value of most experimentally measured quantities have the same approximate value for all states that are within a narrow energy window---hence, if we average over only those states in a given energy window that share the same symmetry as the ground state, they will approximate the results that average over all states, as long as all energy windows are represented in the subspace with the fixed quantum numbers.

The antiferromagnetic case presents a different story. Because the system has much smaller gaps, includes a large number of low-lying states, and has a good deal of frustration, this diabatic evolution does not represent an equilibrium thermal state well. We found it difficult to fit the results to a unique effective temperature and the different experimental quantities were not approximated so well by a thermal distribution. This leads us to conclude that while it is possible that one can use generalizations of adiabatic state preparation to create effective thermal distributions, it does not indicate that such an approach will always work. Instead, it shows that there are some systems, which are the more interesting systems, where this approach is likely to fail and the diabatic evolution is not going to create a thermal distribution. The reason why is that once the system starts to excite from the ground state to higher excited states, it will subsequently show both excitation and de-excitation from those higher states. The net result will be a distribution of states that is not governed by a simple exponential in energy.

\section{Conclusion}

In this work, we examined the difference between the diabatic evolution of a quantum system and states represented by thermal distributions. The goal was to determine whether one could use a modification of the adiabatic state preparation approach to create analogs of thermal states in quantum simulators without attaching them to external baths at fixed temperatures. For concreteness, we chose the transverse-field Ising model as the quantum system, with parameters that are similar to those used in recent experiments. We find that the ferromagnetic case does appear to be able to create near thermal distributions of the Ising model. This was verified by examining different ways to extract the effective temperature, and common observables like the Binder cumulant and the spin structure factor. We also found that this approach does not work as well for the antiferromagnetic case, most likely because of the small energy gaps and the large degeneracy of states at low energy which emerge due to frustration in the model. 

If one could create thermal distributions in quantum emulators, it would open the door to a new class of experiments, where one could engineer the temperature by controlling the speed of the diabatic evolution, and then use these quantum emulators to directly test quantum phenomena in a controlled environment but with equilibrium thermal states. Such studies could provide interesting insight into critical phenomena, especially critical exponents, perhaps allowing them to be directly measured in systems where they are difficult to calculate. Further work could investigate whether there are alternative methods that would improve these results, such as varying the shape of the ramp function for the magnetic field, examining the effect of real phonon creation, or adding other terms like those used in shortcuts to adiabaticity~\cite{shortcuts} that might improve the overlap of the diabatic state with the thermal one.

Our results indicate that this approach might be feasible in ferromagnetic systems.  While these systems might not be the most interesting because they do not have frustration, they could serve as a useful paradigm for this type of study and could allow for a number of interesting benchmarks to be measured which mix in both the quantum and the thermal aspects in a controlled environment. We hope that experimental colleagues will investigate these ideas in the near future.

\section*{Acknowledgments}

ML acknowledges support from the National Science Foundation under grant number DMR-1004268.
JF and BY acknowledge support from the National Science Foundation under grant number PHY-1314295. JF also acknowledges support from the McDevitt bequest at Georgetown University. BY acknowledges support from the Achievement Rewards for College Students Foundation.

\end{document}